\documentclass[12pt]{article}

\usepackage{amsfonts,amssymb, amsmath}
\usepackage[english]{babel}

\def\nn{ \nonumber }
\def\bq{ \begin{equation} }
\def\eq{ \end{equation} }
\def\ben{ \begin{eqnarray} }
\def\en{ \end{eqnarray} }

\newtheorem{prop}{Proposition}

\begin{document}


\title{A note on elliptic coordinates\\ on the Lie algebra $e(3)$.}

\author{
A.V. Tsiganov\\
\it\small
St.Petersburg State University, St.Petersburg, Russia\\
\it\small e--mail: tsiganov@mph.phys.spbu.ru}
\date{}
 \maketitle

\begin{abstract}
We introduce elliptic coordinates on the dual space to the Lie
algebra $e(3)$ and discuss the separability of the Clebsch system in
these variables. The proposed Darboux coordinates  on $e^*(3)$
coincide with the usual elliptic coordinates on the cotangent bundle
of the two-dimensional sphere  at the zero value of the corresponding
Casimir function.
\end{abstract}

The Lie algebra $e(3)=so(3) \oplus {\mathbb R}^3$ of the Lie group of
euclidean motions of $\mathbb R^3$ is a semidirect sum of an algebra
$so(3)$ and an abelian ideal  ${\mathbb R}^3$. For convenience we
shall use the invariant inner product to identify the dual of the Lie
algebra, namely $e^*(3)$, with the Lie algebra $e(3)$.

On the dual space $e^*(3)$ with coordinates $J=(J_1,J_2,J_3)\in
so(3)\simeq {\mathbb R}^3$ and $x=(x_1,x_2,x_3)\in {\mathbb R}^3$ the
Lie-Poisson bracket is defined by
\bq\label{e3} \{J_{i}, J_{j}\}=\varepsilon_{ijk}J_{k}, \quad
\{J_{i}, x_{j}\} =\varepsilon_{ijk}x_{k}, \quad \{x_{i},x_{j}\}=0\,,
\eq
where $\varepsilon_{ijk}$ is the sign of the permutation $(ijk)$ of
$(123)$. The Lie-Poisson bracket (\ref{e3}) is degenerated and has
two Casimir functions
\begin{equation}\label{caz0}
A=| x|^2\equiv\sum_{k=1}^3 x_k^2, \qquad
  B=\langle x, J\rangle\equiv\sum_{k=1}^3 x_kJ_k .
\end{equation}
Here $\langle x,J\rangle$ stands for the standard euclidean scalar
product in $\mathbb R^3$. The generic symplectic leaf
\bq\label{symp-e3}
{\mathcal O}_{ab}= \{x,J\,:~A=a^2,~~ B={b}\}\,,
\eq
is a four-dimensional symplectic  manifolds, which is topologically
equivalent to the cotangent bundle $T^*\mathbb S^2$ of the
two-dimensional sphere $\mathbb S^{2}=\{x\in \mathbb R^3, |x|=a\}$
\cite{nov81}.

If $b=0$, then there exists a symplectic transformation
\bq\label{xp-J}
\rho: (p,x)\to J=p\wedge x,\qquad
J_i=\sum_{j,k=1}^{n=3}\varepsilon_{ijk}\,p_j\,x_k\,,
\eq
which identify $T^*\mathbb S^2\subset T^*\mathbb R^3$ and ${\mathcal
O}_{a0}$. Here $p\wedge x$ means the standard euclidean cross product
in $\mathbb R^3$ and vector $p$ is canonically conjugated to $x$
momenta in $T^*\mathbb R^3$, $\{p_i,x_j\}=\delta_{ij}$, such that
$\langle p,x \rangle=0$.

If $b\neq 0$, the symplectic structure of manifold $\mathcal O_{ab}$
differs from the standard symplectic structure of $T^*\mathbb S^2$ by
a magnetic term proportional to $b$ \cite{nov81}.

The aim of this note  is to describe this magnetic term with the help
of elliptic coordinates on the sphere  $\mathbb S^2$ lifted to the
Darboux variables  on the manifold $e^*(3)$ by $b\neq 0$.

The elliptic coordinates $u_1,u_2$ on $\mathbb S^2$ with parameters
$\alpha_1<\alpha_2<\alpha_3$ is defined as roots of the equation
\bq\label{def-u}
e(\lambda)=\sum_{j=1}^3 \dfrac{x_j^2}{\lambda-\alpha_j}=\dfrac{a^2
(\lambda-u_1)(\lambda-u_2)}{\varphi(\lambda) }=0,
\eq
where  $\varphi(\lambda)=\prod_{j=1}^3(\lambda-\alpha_j)$ and $|x| =
a$, see \cite{kal86}. Like the elliptic coordinates in $\mathbb R^3$,
the elliptic coordinates on $\mathbb S^2$ are also orthogonal and
only locally defined. They take values in the intervals
\bq\label{opr-u}
\alpha_1<u_1<\alpha_2<u_2<\alpha_3.
\eq
By using the Lie-Poisson bracket (\ref{e3}) we can prove that
elliptic coordinates $u_{1,2}$ and variables
\bq\label{def-pi0}
\pi^0_{1,2}=h(u_{1,2}),\qquad
h(\lambda)=\dfrac{1}{2a^2}\sum_{j=1}^3\dfrac{x_j\,(x\wedge
J)_j}{\lambda-\alpha_j}
\eq
satisfy to the following relations
\bq\label{pi0-u}
\{u_1,u_2\}=0,\qquad\{\pi_i^0,u_j\}=\delta_{ij},\qquad
\{\pi_1^0,\pi_2^0\}=\dfrac{ib}{4a}\,\dfrac{u_2-u_1}{\sqrt{\varphi(u_1)\varphi(u_2)}}.
\eq
These relations may be easy obtained by means of the Poisson brackets
between the generating functions $e(\lambda)$ and $h(\mu)$
\ben
\{e(\lambda),e(\mu)\}&=&0,\qquad
\{e(\lambda),h(\mu)\}=-a^{-2}e(\lambda)\,e(\mu)-\dfrac{e(\lambda)-e(\mu)}{\lambda-\mu},\nn\\
\nn\\
\{h(\lambda),h(\mu)\}&=&\dfrac{b}{4a^2}
\dfrac{x_1x_2x_3(\lambda-\mu)(\alpha_1-\alpha_2)(\alpha_2-\alpha_3)(\alpha_3-\alpha_1)}
{\varphi(\lambda)\,\varphi(\mu)}\,.\nn
\en
If $b=0$ relations (\ref{pi0-u}) yield well known fact that variables
$u_{1,2}$ and $\pi_{1,2}^0$ are the Darboux coordinates on the
manifold $T^*\mathbb S\simeq \mathcal O_{a0}$.

The coordinates $u_i$ and the parameters $\alpha_j$ can be subjected
to a simultaneous linear transformation $u_i\to \beta u_i+\gamma$ and
$\alpha_j\to \beta \alpha_j+\gamma$, so it is always possible to
choose
\[\alpha_1=0, \qquad \alpha_2=1, \qquad \alpha_3=\mathtt k^2>1.\]
Using relations (\ref{pi0-u}) and this choice of parameters
$\alpha_j$ we can prove the following
\begin{prop}
If $b\neq 0$,  elliptic coordinates  $u_{1,2}$ (\ref{def-u}) and the
corresponding momenta
\bq\label{def-pi}
\pi_{1,2}=\pi_{1,2}^0-bf_{1,2},\qquad f_{1,2}=
\dfrac{u_{1,2}}{2a\sqrt{\varphi(u_{1,2})}}\,
F\left(\dfrac{\sqrt{\mathtt k^2-u_{2,1} }}{\mathtt k},\dfrac{\mathtt
k}{\mathtt k^2-1}\right),
 \eq
form a complete set of the Darboux variables on the manifold
 $e^*(3)$
\[\{u_1,u_2\}=\{\pi_1,\pi_2\}=0,\qquad\{\pi_i,u_j\}=\delta_{ij}\,, \]
which  are real variables in their domain of definition
(\ref{opr-u}). Here $F(z,k)$ is incomplete elliptic integral of the
first kind, which is  identical to the inverse function of the
elliptic Jacobi function $sn(z,k)$ \cite{as65}.
\end{prop}
\textbf{Proof}: The functions $f_{1,2}$ depend on the coordinates
$u_{1,2}$ and, therefore, we have to verify one relation only
\[
\{\pi_1,\pi_2\}=\{\pi^0_1,\pi^0_2\}-b\{\pi_1^0,f_2\}+b\{\pi_2^0,f_1\}=
\{\pi^0_1,\pi^0_2\}-b\left(\dfrac{\partial f_2}{\partial u_1}-
\dfrac{\partial f_1}{\partial u_2}\right)=0\,.
\]
This relation follows from (\ref{pi0-u}) and properties of the
incomplete elliptic integral of the first kind $F(z,k)$.

So, equations (\ref{def-u}), (\ref{def-pi0}) and (\ref{def-pi})
define elliptic variables $u$ and $\pi$ on $e^*(3)$ as functions on
initial variables $x$ and $J$. The inverse transformation $(u,\pi)\to
(x,J)$ reads as
\bq\label{inv-tr}
x_j=a\sqrt{\dfrac{(\alpha_j-u_1)(\alpha_j-u_2)}{(\alpha_j-\alpha_m)(\alpha_j-\alpha_n)}\,
}, \qquad J_j=a^{-2}\Bigl(bx_j+(z\wedge x)_j\Bigr),
\eq
where $m\neq j\neq n$ and entries of the vector $z$ are given by
\[
z_j=\dfrac{2\,x_j}{u_1-u_2}\left(\dfrac{\varphi(u_1)(\pi_1+bf_1)}{\alpha_j-u_1}
-\dfrac{\varphi(u_2)(\pi_2+bf_2)}{\alpha_j-u_2}\right)\,.
\]
Like the elliptic coordinates in $\mathbb R^n$ and on $\mathbb S^n$
\cite{kal86}, the elliptic variables on $e^*(3)$ may be successfully
exploited in the theory of integrable systems.  For instance,
substituting expressions (\ref{inv-tr}) into the quadratic
hamiltonian
\[
H=\dfrac{1}{2}(J_1^2+J_2^2+J_3^2)+\dfrac12(\alpha_1x_1^2+\alpha_2x_2^2+\alpha_3x_3^2),
\]
associated with the integrable  Clebsch system on $e^*(3)$, one gets
this hamiltonian in terms of the elliptic variables
\[
H=\frac{2}{u_1-u_2}\Bigl(
\varphi(u_1)(\pi_1+bf_1)^2-\varphi(u_2)(\pi_2+bf_2)^2\Bigr)+\frac{b^2}{2a^2}
-\frac{a^2(u_1+u_2-\sum_{j=1}^3 \alpha_j)}2\,.
\]
Here  $f_{1,2}$ are  functions on $u_1$ and $u_2$ and, therefore, the
hamiltonian $H$  belongs to the St\"{a}ckel family of integrals of
motion at $b=0$ only, i.e. for the Neumann system on the sphere
\cite{kal86}. In spite of the fact that $f_{1,2}$ are elliptic functions 
this hamiltonian $H$ may be rewritten as rational
quasi-St\"{a}ckel hamiltonian introduced in \cite{sokmar06}.

Thus we have to fully appreciate that elliptic variables $u_{1,2}$
and $\pi_{1,2}$ cannot be the separation variables for the
Hamilton-Jacobi equation associated with the Clebsch system at $b\neq
0$. The similar result has been obtained in \cite{harl,ts05} by using
velocities
\[\dot{u}_{1,2}=\frac{\mp4\varphi(u_{1,2})(\pi_{1,2}+bf_{1,2})}{u_1-u_2}\]
instead of the momenta $\pi_{1,2}$.  It will be interesting to compare 
this result with results of the  paper \cite{sokmar5} about separability of the
Clebsch system by using the "Kowalevski variables", which in fact coincide
with the elliptic coordinates (\ref{def-u}).

On the other hand, we can substitute Darboux variables $u_{1,2}$ and
$\pi_{1,2}$ into the usual St\"{a}ckel integrals of motion and get a
whole family of integrable systems on the manifold $e^*(3)$, which
are separable in these variables. The main problem in this widely
known Jacobi method is how to single out integrable systems
interesting in physics from this huge family.

The research  was partially supported by  the RFBR grant
 06-01-00140. The author want to thank V.G. Marikhin for valuable
discussions.


\begin{thebibliography}{10}
\bibitem{nov81}
S.P. Novikov,
\newblock{\em Hamiltonian formalism and multi-valued analog of Morse
theory}, Uspekhi Mat. Nauk, \textbf{37}, 3-49, 1982.

\bibitem{kal86}
E. G. Kalnins, {\em Separation of Variables for Riemannian Spaces of
Constant Curvature}, Longman Scientific $\&$ Technical, Essex, 1986.

\bibitem{as65}  M. Abramowitz and I. Stegun,
\newblock{\em Handbook of Mathematical Functions},  Dover
Publications Inc., New York, 1046 p., 1965.




\bibitem{sokmar06}
V.G. Marikhin, V.V. Sokolov,
\newblock{\em On the quasi-St\"{a}ckel hamiltonians},
Uspekhi Mat. Nauk, \textbf{60}(5), 175-176, 2006.



\bibitem{harl}
E.I. Kharlamova,
\newblock{\em O dvigenii tverdogo tela vokrug nepodvignoj
tochki v tsentral'nom n'yutonovskom pole},
\newblock{{Izvestiya Sib. Otd. AN SSSR}, \textbf{6}, 7-17, 1959.}




\bibitem{ts05}
I.V. Komarov, A.V. Tsiganov,
\newblock{\em On a trajectory
isomorphism of the Kowalevski gyrostat and the Clebsch problem},
 J. Phys.A, \textbf{38}, 2917-2927, 2005.


\bibitem{sokmar5}
V.G. Marikhin, V.V. Sokolov, \newblock{\em Separation of variables on
a non-hyperelliptic curve}, Reg. Chaot. Dyn, \textbf{10}(1),
59-70,2005.


\end{thebibliography}
\end{document}